\documentclass[12pt]{article}
\usepackage{a4wide}
\usepackage{amsmath,amssymb,amsfonts}
\usepackage{graphicx,subfigure,epsfig}
\usepackage{xcolor}
\usepackage[T1]{fontenc}
\usepackage{upquote,multirow,slashed}
\usepackage{authblk}
\usepackage{cite,appendix}
\usepackage{empheq}
\usepackage{lipsum}

\definecolor{ForestGreen}{rgb}{0.15,0.70,0.15}
\usepackage[colorlinks=true]{hyperref}
\hypersetup{
 linktocpage=true,
 citecolor=ForestGreen,
 linkcolor=blue,
 urlcolor=blue}
\date{\vspace{-5ex}}

\newcommand{\NOdisplay}[1]{ }

\def\MSbar{\overline{\mathrm{MS}}}
\def\TR{{\displaystyle \mathrm{T}_{F}}}
\def\gJJ{\gamma_{s}}
\def\gFJ{\gamma_{{\scriptscriptstyle FJ}}}
\def\gFF{\gamma_{{\scriptscriptstyle F\tilde{F}}}}

\begin{document}

% ******************************* 
% Title & details of the authors
% *******************************

\title{\vspace{-2cm}
{\small \hfill TTK-22-01, P3H-22-001 \\} 
\vspace{2cm}
\textbf{The $\overline{\mathrm{MS}}$ renormalization constant of the singlet axial current operator at $\mathcal{O}(\alpha_s^5)$ in QCD}}

\author[a,b]{Long Chen\footnote{\textit{E-mail}: 
longchen@sdu.edu.cn}}
\author[b]{Micha\l{} Czakon\footnote{\textit{E-mail}: mczakon@physik.rwth-aachen.de}}
\affil[a]{School of Physics, Shandong University, Jinan, Shandong 250100, China}
\affil[b]{Institut f\"ur Theoretische Teilchenphysik und Kosmologie, RWTH Aachen University, Sommerfeldstr.~16, 52056 Aachen, Germany}

\maketitle

\noindent\rule{\textwidth}{.5pt}
\begin{abstract}
We provide the $\overline{\mathrm{MS}}$ factor of the renormalization constant of the singlet axial-current operator in dimensional regularization at $\mathcal{O}(\alpha_s^5)$ in perturbative Quantum Chromodynamics (QCD). The result is obtained from a formula derived using the Adler-Bell-Jackiw equation in terms of renormalized operators.
The required input consists of the complete four-loop results for operator renormalization constants obtained by explicit diagrammatic computation in a previous publication.
\end{abstract}
\noindent\rule{\textwidth}{.5pt}

\thispagestyle{empty}

\clearpage

%\noindent\rule{\textwidth}{.5pt}
%{
%\hypersetup{linkcolor=blue}
%\tableofcontents
%}
%\noindent\rule{\textwidth}{.5pt}

%\thispagestyle{empty}
%\clearpage
%\vspace{1cm}

\allowdisplaybreaks

As a consequence of the famous Adler-Bell-Jackiw (ABJ) anomaly~\cite{Adler:1969gk,Bell:1969ts}, the singlet axial-current operator requires a non-trivial renormalization~\cite{Adler:1969gk,Chanowitz:1979zu,Trueman:1979en,Kodaira:1979pa,Gottlieb:1979ix,Espriu:1982bw,Collins:1984xc,Bos:1992nd,Larin:1991tj,Larin:1993tq}.
The ultraviolet (UV) renormalized singlet axial-current operator thus has a non-zero anomalous dimension, regardless of the particular regularization scheme used in the treatment of the $\gamma_5$ matrix. 
One interesting implication of this non-trivial point in practical physical applications is the appearance of non-decoupling heavy-quark-mass logarithms~\cite{Collins:1978wz,Chetyrkin:1993hk} in the axial quark form factors, which are present in an anomaly-free theory such as the Standard Model.
These mass logarithms can be resummed by the renormalization group (RG) method where the anomalous dimension involved is precisely that of the renormalized singlet axial current~\cite{Chetyrkin:1993jm,Chetyrkin:1993ug,Larin:1993ju,Larin:1994va,Chen:2021rft}.
Dimensional regularization (DR)~\cite{tHooft:1972tcz,Bollini:1972ui}  offers a very efficient framework for calculating the renormalization constant of the axial-current operator, as well as the related non-vanishing anomalous dimension. 
After the initial studies of the non-trivial renormalization needed for a flavor singlet axial current~\cite{Adler:1969gk,Chanowitz:1979zu,Trueman:1979en,Kodaira:1979pa,Gottlieb:1979ix,Espriu:1982bw,Collins:1984xc,Bos:1992nd}, especially a pure $\MSbar$ renormalization being insufficient, its result was previously determined in DR up to $\mathcal{O}(\alpha_s^3)$ in QCD in refs.~\cite{Larin:1993tq,Ahmed:2021spj}, with the pure $\overline{\mathrm{MS}}$ part known to $\mathcal{O}(\alpha_s^4)$~\cite{Larin:1997qq}.\footnote{The $\mathcal{O}(\alpha_s^2)$ results were determined and discussed also in the calculation of polarized splitting functions, e.g.~refs.\cite{Matiounine:1998re,Vogt:2008yw,Moch:2014sna,Behring:2019tus}.}
The complete result for this quantity at four-loop order in QCD has recently been obtained in ref.~\cite{Chen:2021gxv} by means of an off-shell axial Ward-Takahashi identity~\cite{Adler:1969gk} (with non-vanishing contact terms) using a particular variant of the non-anticommutating $\gamma_5$ definition~\cite{Larin:1991tj,Larin:1993tq}.

In this letter, we derive the result for the $\overline{\mathrm{MS}}$ factor, $Z_s^{ms}$, i.e.\ the pure-pole contribution (principal part of the Laurent expansion) with respect to the DR parameter $\epsilon$ corresponding to spacetime dimension $D = 4-2\epsilon$, of the renormalization constant of the singlet axial-current operator at $\mathcal{O}(\alpha_s^5)$ in QCD. To this end, we exploit a formula following from the all-order axial anomaly equation~\cite{Adler:1969gk,Adler:1969er}.
Our result belongs to the small number of renormalization constants currently known at five-loop order accuracy, albeit it does not require performing an actual five-loop computation.
As will be shown explicitly below, the anomalous dimension of the fully renormalized singlet axial-current at $\mathcal{O}(\alpha_s^5)$ in the 4-dimensional limit does not receive contributions from the (unknown) $\mathcal{O}(\alpha_s^5)$ result for the finite factor of the axial-current renormalization constant $Z_s^f$, which is introduced on top of the pure $\overline{\mathrm{MS}}$ part to restore the correct form of the anomalous axial Ward identity (see below), although it does involve the $\mathcal{O}(\alpha_s^4)$ result. 
On the other hand, the same anomalous dimension can be determined from the four-loop result for the r.h.s.~of the properly renormalized axial-anomaly equation.
After carefully isolating the $\mathcal{O}(\alpha_s^5)$ result for the anomalous dimension of $Z_s^{ms}$, this pure $\overline{\mathrm{MS}}$ renormalization constant can then be uniquely reconstructed.
~\\

We follow, here and in the remainder of this letter, the notations and conventions of our previous work~\cite{Chen:2021gxv}.
For the sake of readers' convenience, we briefly recapitulate the essentials before we dive into the main work of this publication.
The all-order axial-anomaly equation~\cite{Adler:1969gk,Adler:1969er} in QCD with $n_f$ massless quarks, expressed in terms of renormalized local composite operators, reads: %
\begin{eqnarray} 
\label{eq:ABJanomalyEQ}
\big[\partial_{\mu} J^{\mu}_{5,s} \big]_{R} = a_s\, n_f\, \TR \,  \big[F \tilde{F} \big]_{R}\,,
\end{eqnarray}
where the subscript $R$ at a square bracket denotes operator renormalization, $a_s \equiv \frac{\alpha_s}{4 \pi} = \frac{g_s^2}{16 \pi^2}$ is a shorthand notation for the QCD coupling, while $\TR=1/2$. The renormalization of the operators $\partial_{\mu} J^{\mu}_{5,s}$ and $F \tilde{F}$ can be arranged into the following matrix form~\cite{Adler:1969gk,Trueman:1979en,Espriu:1982bw,Breitenlohner:1983pi}: %
\begin{eqnarray}
\label{eq:Zsmatrix}
\begin{pmatrix}
\big[\partial_{\mu} J^{\mu}_{5,s} \big]_{R}\\
\big[F \tilde{F}\big]_{R}
\end{pmatrix}
= 
\begin{pmatrix}
Z_{s} &  0 \\
Z_{FJ} &  Z_{F\tilde{F}}
\end{pmatrix}
\cdot 
\begin{pmatrix}
\big[\partial_{\mu} J^{\mu}_{5,s} \big]_{B}\\
\big[F \tilde{F}\big]_{B}
\end{pmatrix}\,,
\end{eqnarray}
where the subscript $B$ at a square bracket indicates that all fields are bare. The singlet axial-current is defined as $\big[ J^{\mu}_{5,s}\big]_{B} \equiv \sum_{i=1}^{n_f} \, \bar{\psi}^{B}_{i}  \, \gamma^{\mu}\gamma_5 \, \psi^{B}_i$ with a non-anticommuting $\gamma_5$, while $F \tilde{F} \equiv  - \epsilon^{\mu\nu\rho\sigma} F^a_{\mu\nu} F^a_{\rho\sigma} \equiv \epsilon_{\mu\nu\rho\sigma} F^a_{\mu\nu} F^a_{\rho\sigma}$ denotes the contraction of the field-strength tensor and its dual, with $F^a_{\mu\nu} \equiv \partial_{\mu} A_{\nu}^{a} - \partial_{\nu} A_{\mu}^{a} + g_s \,  f^{abc} A_{\mu}^{b} A_{\nu}^{c}$, $A_\mu^a$ the gluon field, and $f^{abc}$ the non-abelian color group structure constants. 
Being the only parity-odd gauge-invariant current operator (with a canonical mass dimension 3) in the theory, the singlet axial-current $J^{\mu}_{5,s}$ renormalizes multiplicatively, however, the renormalization of the axial-anomaly operator $F \tilde{F}$, which can be written as the divergence of the gauge-non-invariant Chern-Simons current, is not strictly multiplicative but involves mixing with the divergence of $J^{\mu}_{5,s}$~\cite{Adler:1969gk,Espriu:1982bw,Breitenlohner:1983pi}.
From this follows the lower-triangular form of the mixing matrix in eq.(\ref{eq:Zsmatrix}).
Accordingly, the matrix of anomalous dimensions of the two renormalized operators is defined by: %
\begin{eqnarray}
\label{eq:AMDmatrix}
\frac{\mathrm{d}}{\mathrm{d}\, \ln \mu^2}\,
\begin{pmatrix}
\big[\partial_{\mu} J^{\mu}_{5,s} \big]_{R}\\
\big[F \tilde{F}\big]_{R}
\end{pmatrix}
= 
\begin{pmatrix}
\gJJ &  0 \\
\gFJ &  \gFF
\end{pmatrix}
\cdot 
\begin{pmatrix}
\big[\partial_{\mu} J^{\mu}_{5,s} \big]_{R}\\
\big[F \tilde{F}\big]_{R}
\end{pmatrix} \, ,
\end{eqnarray}
with $\mu$ the mass scale in dimensional regularization.
To be explicit, we have for these anomalous dimensions: % 
\begin{eqnarray}
\label{eq:AMDinZs}
\gJJ &=& \frac{\mathrm{d}\, \ln Z_{s}}{\mathrm{d}\, \ln \mu^2}\,, \quad
\gFF = \frac{\mathrm{d}\, \ln Z_{F\tilde{F}}}{\mathrm{d}\, \ln \mu^2}\,, \nonumber\\
\gFJ &=& \frac{1}{Z_s}\frac{\mathrm{d}\, Z_{FJ}}{\mathrm{d}\, \ln \mu^2}
- \frac{Z_{FJ}}{Z_s} \frac{\mathrm{d}\, \ln Z_{F\tilde{F}}}{\mathrm{d}\, \ln \mu^2}\,. 
\end{eqnarray}
A few comments are in order.
$Z_{F\tilde{F}}$ and $Z_{FJ}$ are pure $\MSbar$ renormalization constants.
In particular, $Z_{F\tilde{F}}$ in $\MSbar$ scheme is equal to the $\MSbar$ QCD-coupling renormalization constant $Z_{a_s}$, as verified explicitly to $\mathcal{O}(\alpha_s^4)$ in ref.~\cite{Ahmed:2021spj} and proven recently in ref.~\cite{Luscher:2021bog}.
Consequently, $\gFF$ is equal to the negative of the QCD beta function defined as $\beta \equiv - \frac{\mathrm{d} \ln Z_{a_s}}{\mathrm{d} \ln \mu^2}$.
In contrast, the multiplicative renormalization constant $Z_s$ of the singlet axial-current is not pure $\MSbar$, and is conveniently parameterized as the product of a pure $\MSbar$-renormalization factor $Z^{ms}_{s}$ and an additional finite, by definition $\epsilon$-independent, renormalization factor $Z^{f}_{s}$, namely $Z_{s} \equiv Z^{f}_{s} \, Z^{ms}_{s}$.
The latter is needed to restore the correct form of the anomalous axial Ward identity following from inserting eq.~(\ref{eq:ABJanomalyEQ}) into (on-shell) S-matrix elements.\footnote{
For instance, when considering the matrix element of $J^{\mu}_{5,s}$ with an external momentum insertion $q$ between a pair of external quark states with on-shell momenta $p$ and $p'$ in the momentum space, the anomalous axial Ward identity in this case reads~\cite{Adler:1969gk} $q_{\mu}\, \mathrm{\Gamma}^{\mu}_{5,s}(p',p) = -a_s\, n_f\, \TR \,\mathrm{\Lambda}(p',p)$ where $\mathrm{\Gamma}^{\mu}_{5,s}(p',p)$ and $\mathrm{\Lambda}(p',p)$ denote, respectively, the amputated one-particle irreducible 3-point vertex function computed using the renormalized axial current and anomaly operator in eq.~(\ref{eq:ABJanomalyEQ}).
In case of massive quark fields, there will be an additional contribution generated by the classically-expected mass term.}
Still, since $Z_s$, $Z_{F\tilde{F}}$ and $Z_{FJ}$ all have no explicit dependence on the scale $\mu$, the logarithmic derivative $\frac{\mathrm{d}\,}{\mathrm{d}\, \ln \mu^2}$ in eq.~(\ref{eq:AMDinZs}) can be rewritten in terms of the derivative w.r.t.~$a_s$ with the help of the RG equation of $a_s$ in $D$ dimensions, which reads %
\begin{eqnarray}
\label{eq:beta}
\frac{\mathrm{d} \ln a_s}{\mathrm{d} \ln \mu^2} = -\epsilon -\frac{\mathrm{d} \ln Z_{a_s}}{\mathrm{d} \ln \mu^2} \equiv -\epsilon + \beta \,.
\end{eqnarray}

Due to the appearance of the finite non-$\MSbar$ renormalization factor $Z^{f}_{s}$, the full anomalous dimension $\gJJ = \frac{\mathrm{d}\, \ln Z_{s}}{\mathrm{d}\, \ln \mu^2}$ contains terms suppressed by powers of $\epsilon$ in $D$ dimensions. In fact, these extra terms are strictly linear in $\epsilon$ to all orders. 
To see this, we write: %
\begin{eqnarray}
\label{eq:gJdec}
\gJJ &=& \frac{\mathrm{d}\, \ln Z_{s}}{\mathrm{d}\, \ln \mu^2}\, = \frac{\mathrm{d}\, \ln  Z^{ms}_{s} }{\mathrm{d}\, \ln \mu^2}\, + \, \frac{\mathrm{d}\, \ln Z^{f}_{s} }{\mathrm{d}\, \ln \mu^2}\,\nonumber\\
&\equiv&  \gamma_{s}^{ms} \,+\,  \gamma_{s}^{f}\,,
\end{eqnarray} %
where we have introduced $\gamma_{s}^{ms} \equiv  \frac{\mathrm{d}\, \ln  Z^{ms}_{s} }{\mathrm{d}\, \ln \mu^2}$ and $\gamma_{s}^{f} \equiv \frac{\mathrm{d}\, \ln Z^{f}_{s} }{\mathrm{d}\, \ln \mu^2}$.
Apparently $\gamma_{s}^{ms}$ is independent of $\epsilon$, just like $\gFF=-\beta$ and $\gFJ$, as $Z^{ms}_{s}$ is a $\MSbar$ renormalization constant.
However, by eq.~(\ref{eq:beta}), $\gamma_{s}^{f}$ has a linear dependence in $\epsilon$: %
\begin{eqnarray}
\label{eq:gJfdec}
\gamma_{s}^{f} & \equiv& \frac{\mathrm{d}\, \ln Z^{f}_{s} }{\mathrm{d}\, \ln \mu^2} = \frac{\mathrm{d}\, \ln a_{s} }{\mathrm{d}\, \ln \mu^2}\,\frac{\mathrm{d}\, \ln Z^{f}_{s} }{\mathrm{d}\, \ln a_s}   \nonumber\\ 
&=&  \beta\, \frac{\mathrm{d}\, \ln Z^{f}_{s} }{\mathrm{d}\, \ln a_s} \,-\, \epsilon\, \frac{\mathrm{d}\, \ln Z^{f}_{s} }{\mathrm{d}\, \ln a_s}\,,
\end{eqnarray}
where we have used the fact that the finite $Z^{f}_{s}$ has no explicit dependence on either $\mu$ or $\epsilon$\footnote{One recalls that \textit{by definition} the $Z^{f}_{s}$ is determined via demanding the correct form of the axial Ward identity following from the all-order axial anomaly equation hold in the 4-dimensional limit (i.e.~$\epsilon=0$)\cite{Larin:1993tq,Bos:1992nd,Chetyrkin:1993hk}, because after all there is no axial anomaly in a generic D dimensions.}.
We thus see clearly that the $\epsilon$-independent part of $\gamma_{s}^{f}$ is proportional to its $\epsilon$-linear part by exactly a factor $-\beta$. 
On the other hand, subjecting both sides of the axial anomaly equation~\eqref{eq:ABJanomalyEQ} to the logarithmic derivative $\frac{\mathrm{d}\, }{\mathrm{d}\, \ln \mu^2}$,  
one obtains the relation~\cite{Breitenlohner:1983pi,Bos:1992nd,Larin:1993tq,Ahmed:2021spj,Luscher:2021bog}: %
\begin{eqnarray}
\label{eq:AMDsEQ}
\bar{\gamma}_{s} \equiv \gJJ|_{\epsilon = 0} = a_s\,  n_f \, \TR \, \gFJ \,,
\end{eqnarray}%
with the aid of the equality $\gFF=-\beta$.
Now, combining eq.~(\ref{eq:gJdec}), eq.~(\ref{eq:gJfdec}) and eq.~(\ref{eq:AMDsEQ}), we arrive at our master formula for reconstructing  $Z^{ms}_{s}$: % 
\begin{eqnarray}
\label{eq:gJms}
\gamma_{s}^{ms} & \equiv& \frac{\mathrm{d}\, \ln Z^{ms}_{s} }{\mathrm{d}\, \ln \mu^2} \nonumber\\
&=&  a_s\,  n_f \, \TR \, \gFJ \,-\, 
\beta\, \frac{\mathrm{d}\, \ln Z^{f}_{s} }{\mathrm{d}\, \ln a_s} \,.
\end{eqnarray} %
A key feature of this formula, which makes it advantageous for determining $Z^{ms}_{s}$, is that the finite $\gFJ$ and $Z^{f}_{s}$ appearing on its r.h.s.~have coefficients whose perturbative expansions start from $\mathcal{O}(a_s)$.
Consequently, to obtain the perturbative result for $\gamma_{s}^{ms}$ on the l.h.s., and hence $Z^{ms}_{s}$, at $\mathcal{O}(a_s^N)$ one only needs to know $\gFJ$ and $Z^{f}_{s}$ to order $\mathcal{O}(a_s^{N-1})$.\footnote{In the case of a non-singlet axial current, a simpler formula with similar features holds which can be obtained by setting the $\gamma_{FJ}$ term in eq.~(\ref{eq:gJms}) to zero, because a physical non-singlet axial current should be non-anomalous. %and hence the anomalous dimension $\gamma_{ns}^{ms}$ is completely artificial. 
Consequently, the $\mathcal{O}(a_s^5)$ result for the $\MSbar$ renormalization constant of a non-singlet axial current with a non-anticommuting $\gamma_5$ can also be derived from the four-loop result in eq.(4.2) provided in ref.~\cite{Chen:2021gxv} (see the supplementary file).
We thank K.~G.~Chetyrkin for pointing this out to us explicitly.}
Formula~(\ref{eq:gJms}) also explains why the perturbative correction to $Z^{ms}_{s}$ starts at $\mathcal{O}(a_s^2)$ with a simple pole, i.e.~there is no $\mathcal{O}(a_s)$ term.

With both $Z^{f}_{s}$ and $\gFJ$ determined to $\mathcal{O}(a_s^4)$ in ref.~\cite{Chen:2021gxv}, using the formula~(\ref{eq:gJms}) the five-loop order result for $\gamma_{s}^{ms} = \sum_{n=2}^{\infty} a_s^n \, \gamma^{ms}_{s,[n]}$ reads: 
\begin{eqnarray}
\label{eq:gms5}
\gamma^{ms}_{s,[2]} &=& 
C_A C_F \, \left(-\frac{44}{3}\right) \,+\, C_F n_f \, \left(-\frac{10}{3}\right)
\,,\nonumber\\
\gamma^{ms}_{s,[3]} &=& 
C_A^2 C_F \, \left(-\frac{3578}{27}\right)\,+\,C_A C_F^2 \, \left(\frac{308}{3}\right)\,+\,C_A C_F n_f \, \left(-\frac{149}{27}\right)
\nonumber\\&+&
C_F^2 n_f \, \left(\frac{22}{3}\right)\,+\,C_F n_f^2 \, \left(-\frac{26}{27}\right)
\,,\nonumber\\
\gamma^{ms}_{s,[4]} &=& 
C_A^3 C_F \, \left(616 \zeta _3-\frac{36607}{27}\right)\,+\,C_A^2 C_F^2 \, \left(\frac{58618}{27}-1760 \zeta _3\right)
\nonumber\\&+&
C_A^2 C_F n_f \, \left(\frac{874 \zeta _3}{3}+\frac{15593}{162}\right)
\,+\,
C_A C_F^3 \, \left(1056 \zeta _3-\frac{1870}{3}\right)
\nonumber\\&+&
C_A C_F^2 n_f \, \left(\frac{1897}{27}-\frac{184 \zeta _3}{3}\right)\,+\,C_A C_F n_f^2 \, \left(\frac{212 \zeta _3}{3}+\frac{124}{81}\right)
\nonumber\\&+&
C_F^3 n_f \, \left(-192 \zeta _3-\frac{29}{3}\right)
\,+\,
C_F^2 n_f^2 \, \left(\frac{1702}{27}-\frac{224 \zeta _3}{3}\right)
\,+\,
C_F n_f^3 \, \left(\frac{70}{81}\right)
\,,\nonumber\\
\gamma^{ms}_{s,[5]} &=& 
C_A^4 C_F \, \left(\frac{617720 \zeta _3}{81}-\frac{181280 \zeta _5}{9}-\frac{1694 \pi ^4}{45}-\frac{6098299}{486}\right)
\nonumber\\&+&
C_A^3 C_F^2 \, \left(-\frac{75544 \zeta _3}{3}+\frac{748880 \zeta _5}{9}+\frac{968 \pi ^4}{9}+\frac{6674474}{243}\right)
\nonumber\\&+&
C_A^3 C_F n_f \, \left(\frac{330784 \zeta _3}{81}+\frac{36755 \zeta _5}{9}-\frac{2959 \pi ^4}{270}+\frac{2182747}{972}\right)
\nonumber\\&+&
C_A^2 C_F^3 \, \left(\frac{91504 \zeta _3}{9}-\frac{297440 \zeta _5}{3}-\frac{968 \pi ^4}{15}-\frac{535657}{27}\right)
\nonumber\\&+&
C_A^2 C_F^2 n_f \, \left(\frac{19750 \zeta _3}{27}-\frac{184120 \zeta _5}{9}-\frac{2134 \pi ^4}{135}-\frac{206648}{243}\right)
\nonumber\\&+&
C_A^2 C_F n_f^2 \, \left(\frac{8722 \zeta _3}{9}-\frac{3110 \zeta _5}{3}-\frac{146 \pi ^4}{135}+\frac{14803}{243}\right)
\nonumber\\&+&
C_A C_F^4 \, \left(13904 \zeta _3+\frac{80960 \zeta _5}{3}+\frac{43450}{9}\right)
\,+\,
C_F n_f^4 \, \left(2-\frac{160 \zeta _3}{81}\right)
\nonumber\\&+&
C_A C_F^3 n_f \, \left(-1436 \zeta _3+21840 \zeta _5+\frac{352 \pi ^4}{15}-\frac{8131}{18}\right)
\nonumber\\&+&
C_A C_F^2 n_f^2 \, \left(-\frac{25112 \zeta _3}{27}+\frac{80 \zeta _5}{9}+\frac{524 \pi ^4}{135}+\frac{262457}{486}\right)
\nonumber\\&+&
C_A C_F n_f^3 \, \left(-\frac{7700 \zeta _3}{81}+\frac{106 \pi ^4}{135}-\frac{181}{54}\right)
\,+\,
C_3 C_F \, \left(\frac{320}{9}-\frac{2816 \zeta _3}{3}\right)
\nonumber\\&+&
C_F^3 n_f^2 \, \left(-\frac{3016 \zeta _3}{9}+\frac{3680 \zeta _5}{3}-\frac{32 \pi ^4}{15}-\frac{11915}{27}\right)
\nonumber\\&+&
C_1 C_A C_F n_f \, \left(3872 \zeta _3-\frac{22880 \zeta _5}{3}-\frac{14080}{9}\right)
\nonumber\\&+&
C_2 C_A C_F \, \left(\frac{66880 \zeta _3}{3}+28160 \zeta _5-\frac{1408}{9}\right)
\nonumber\\&+&
C_F^4 n_f \, \left(-3104 \zeta _3-\frac{14720 \zeta _5}{3}-\frac{5335}{9}\right)
\nonumber\\&+&
C_F^2 n_f^3 \, \left(\frac{1024 \zeta _3}{27}-\frac{112 \pi ^4}{135}+\frac{16337}{243}\right)
\nonumber\\&+&
C_1 C_F n_f^2 \, \left(-\frac{4160 \zeta _3}{3}+\frac{4160 \zeta _5}{3}+\frac{1792}{3}\right)
\nonumber\\&+&
C_2 C_F n_f \, \left(-\frac{5504 \zeta _3}{3}-5120 \zeta _5-\frac{1792}{9}\right)
\,.
\end{eqnarray}
%
%The result for $Z^{ms}_{s}$ up to $\mathcal{O}(a_s^4)$ can be found in ref.~\cite{Chen:2021gxv}.
The explicit expression for $Z^{ms}_{s} = 1 + \sum_{n=1}^{\infty} a_s^n \, Z^{ms}_{s,[n]}$ reconstructed from eq.~(\ref{eq:gms5}) up to order $\mathcal{O}(\alpha_s^5)$ can be found in the supplementary file.
The definition of the quadratic Casimir color constants is as usual: $C_A = N_c \,, \, C_F = (N_c^2 - 1)/(2 N_c) \,$ with $N_c =3$ in QCD and we have set the color-trace normalization factor to its value $\TR = {1}/{2}$.
Eq.~(\ref{eq:gms5}) implies that the $\mathcal{O}(\alpha_s^5)$ result for $\gamma_{s}^{ms}$ involves three non-quadratic color constants.
These additional color constants appearing in eq.~(\ref{eq:gms5}) can be expressed in terms of contractions of symmetric color tensors\footnote{The symmetric tensor $d_F^{abcd}$ is defined by the color trace $\frac{1}{6} \mbox{Tr}$~$\big( T^a T^b T^c T^d + T^a T^b T^d T^c + T^a T^c T^b T^d + T^a T^c T^d T^b + T^a T^d T^b T^c + T^a T^d T^c T^b \big)$ with $T^a$ the generators of the fundamental representation of the SU($N_c$) group, and similarly $d_A^{abcd}$ for the adjoint representation.}:
\begin{eqnarray}
C_1 &\equiv& \frac{d_F^{abcd} d_F^{abcd}}{N_c^2-1} = \frac{N_c^4-6 N_c^2+18}{96 N_c^2} 
\,,\,\nonumber\\
C_2 &\equiv& \frac{d_F^{abcd} d_A^{abcd}}{N_c^2-1} = \frac{N_c \left(N_c^2+6\right)}{48}
\,,\,\nonumber\\
C_3 &\equiv& \frac{d_A^{abcd} d_A^{abcd}}{N_c^2-1} = \frac{N_c^2 \left(N_c^2+36\right)}{24} 
\,.
\end{eqnarray} %
Hence, due to the $Z_s^f$ term in eq.~(\ref{eq:gJms}), $\gamma_{s}^{ms}$ has a more complicated structure compared to that of $\bar{\gamma}_{s} = \sum_{n=2}^{\infty} a_s^n \, \bar{\gamma}_{s,[n]}$, 
which reads: 
\begin{eqnarray}
\label{eq:gs5}
\bar{\gamma}_{s,[2]} &=& C_F n_f \, \big(-6\big)
\,,\nonumber\\
\bar{\gamma}_{s,[3]} &=& 
C_A C_F n_f \, \left(-\frac{142}{3}\right) + C_F^2 n_f \, \left(18\right) + C_F n_f^2 \, \left(\frac{4}{3}\right)
\,,\nonumber\\
\bar{\gamma}_{s,[4]} &=&
C_A^2 C_F n_f \, \left(-\frac{1607}{6}\right) + C_A C_F^2 n_f \, \left(\frac{461}{2}\right) + C_A C_F n_f^2 \, \left(144 \zeta _3-\frac{82}{3}\right) 
\nonumber\\&+&
C_F^3 n_f \, \left(-63\right) + C_F^2 n_f^2 \, \left(107-144 \zeta _3\right) + C_F n_f^3 \, \left(\frac{26}{3}\right)
\,,\nonumber\\
\bar{\gamma}_{s,[5]} &=&
C_A^3 C_F n_f \, \left(-\frac{17080 \zeta _3}{9}+3520 \zeta _5-\frac{220280}{81}\right) \,+\,
C_F^4 n_f \, \big(9\big)
\nonumber\\&+&
C_A^2 C_F^2 n_f \, \left(1936 \zeta _3-3520 \zeta _5+\frac{230284}{81}\right) \,+\,  C_A C_F^3 n_f \, \left(-\frac{704 \zeta _3}{3}-\frac{6298}{9}\right) 
\nonumber\\&+&
C_A^2 C_F n_f^2 \, \left(\frac{10120 \zeta _3}{3}-1600 \zeta _5-\frac{176 \pi ^4}{15}+\frac{35143}{81}\right)
\nonumber\\&+&
C_A C_F^2 n_f^2 \, \left(-\frac{8128 \zeta _3}{3}-320 \zeta _5+\frac{176 \pi ^4}{15}+\frac{75334}{81}\right)
\nonumber\\&+&
C_A C_F n_f^3 \, \left(-\frac{544 \zeta _3}{3}+\frac{32 \pi ^4}{15}-\frac{3908}{81}\right) \,+\, C_F^2 n_f^3 \, \left(\frac{544 \zeta _3}{3}-\frac{32 \pi ^4}{15}+\frac{8266}{81}\right)
\nonumber\\&+&
C_F^3 n_f^2 \, \left(-\frac{1792 \zeta _3}{3}+1920 \zeta _5-\frac{7136}{9}\right) \,+\, C_F n_f^4 \, \left(\frac{40}{3}-\frac{128 \zeta _3}{9}\right)\,,
\end{eqnarray}
where there appear only $C_A,\,C_F$ up to $\mathcal{O}(\alpha_s^5)$.

Had one proceeded as we propose in the present letter, namely using formula~(\ref{eq:gJms}), the four-loop result for $Z_s^{ms}$, provided in ref.~\cite{Larin:1997qq,Chen:2021gxv}, could have been obtained along with the computations performed in refs.~\cite{Larin:1991tj,Larin:1993tq}.
~\\

One immediate application of the anomalous dimension~(\ref{eq:gms5}), as well as (\ref{eq:gs5}), is the determination of the explicit logarithmic part of the Wilson coefficient that encodes the non-decoupling heavy-quark-mass logarithms in a low-energy effective Lagrangian~\cite{Collins:1978wz,Chetyrkin:1993hk,Chetyrkin:1993jm,Chetyrkin:1993ug,Larin:1993ju,Larin:1994va,Chen:2021rft} at $\mathcal{O}(\alpha_s^5)$.
As expected, as long as one is only concerned with matrix elements of a non-anomalous non-singlet current, which may involve an anomaly-free combination of singlet-type diagrams, it is not even necessary to explicitly renormalize the singlet diagrams. 
The final result for a non-singlet quantity should be completely independent of the particular renormalization prescription one may employ for the singlet set of diagrams, whose effect cancels completely in the non-anomalous combination, see, e.g.~, the discussions in refs.~\cite{Chetyrkin:1993hk,Chen:2021rft}. 
However, for the sake of revealing the structure, and performing resummation, of the non-decoupling heavy-quark-mass logarithms, it is very useful to introduce the renormalization of the singlet set of diagrams, and to this end different working renormalization conventions exist in literature, such as Larin's scheme~\cite{Larin:1991tj,Larin:1993tq} and Chetyrkin's scheme~\cite{Chetyrkin:1993hk}.

We note that one can also combine the ideas from ref.~\cite{Larin:1993tq} and~\cite{Bos:1992nd,Chetyrkin:1993hk}, to easily end up with a renormalized form of the ABJ equation where both sides are manifestly RG-invariant~\cite{Zoller:2013ixa,Luscher:2021bog}. 
To this end, one can simply multiply the following additional finite renormalization constant,
$Z_{ext}(a_s) \equiv \hat{P}\mathrm{exp}\big(\int_{0}^{a_s} \frac{-\bar{\gamma}_s(a)}{\beta(a)} \, \frac{\mathrm{d} a}{a} \big)$ with $\bar{\gamma}_s(a)$ defined in eq.~(\ref{eq:AMDsEQ}), onto both sides of the eq.~(\ref{eq:ABJanomalyEQ}) renormalized already in Larin's convention.
Since $\bar{\gamma}_s(a)$ is $\mathcal{O}(a^2_s)$, $Z_{ext}(a_s)$ has a Taylor power series expansion in $a_s$.
This leads to a new set of $Z_s$, $Z_{F\tilde{F}}$ and $Z_{FJ}$ obtained from the old ones by multiplying $Z_{ext}$, all of which are no longer pure $\MSbar$ renormalization constants; in particular the new $Z_{F\tilde{F}}$ is no longer equal to $Z_{a_s}$ in this special renormalization scheme.
%Below we refer to this choice as the RG-invariant (RGI) scheme for the renormalization of the singlet axial current operator. 

As for the definition of the Wilson coefficient, we follow in particular the conventions of ref.~\cite{Chen:2021rft} for the axial-current coupling to the Z-boson field, $\text{Z}_{\mu}$, in the renormalized low-energy effective Lagrangian:
\begin{equation}
\begin{split}
\label{eq:Leff}
\delta \mathcal{L}^{R}_{\mathrm{eff}} = \mu^\epsilon \Big(& Z_{ns}\,
\sum_{i=1}^{n_f} a_i \,  \bar{\psi}^{B}_{i}  \, \gamma^{\mu}\gamma_5 \, \psi^{B}_i \,+\,
a_b \, Z_{ps} \, \big[ J^{\mu}_{5,s} \big]_B \nonumber\\
\,&+\, a_t\, C_w(a_s, \mu/m_t)\, \big(Z_{ns} + n_f\,Z_{ps} \big) \, \big[ J^{\mu}_{5,s} \big]_B \Big) \text{Z}_{\mu} \,,
\end{split}
\end{equation} %
where $C_w(a_s,\mu/m_t)$ with $m_t$ the on-shell top-quark mass, is the Wilson coefficient in the renormalization scheme implied by the chosen renormalization constants. Furthermore, $a_i$ denotes the axial electroweak coupling of the quark $i$, $a_{d,s,b} = -a_{u,c,t}$. The renormalization constant $Z_{ps} \equiv \frac{1}{n_f} \big(Z_s - Z_{ns} \big)$ is the difference between the singlet and non-singlet axial-current renormalization constants~\cite{Chetyrkin:1993hk,Chetyrkin:1993jm,Chetyrkin:1993ug,Rittinger:2012the,Gehrmann:2021ahy,Chen:2021rft,Chen:2021gxv,Chen:2022aqw}, further normalized to the case of a single quark flavor with axial coupling. 
It is necessary to renormalize diagrams containing a closed $b$-quark loop coupling to the Z-boson. The last term corresponds to the effect of a closed top-quark loop and contains a renormalized singlet current, since $Z_{ns} + n_f\,Z_{ps} = Z_s$. 
(Note that this includes the diagrams with closed top-quark loops but with the Z-boson coupled to a closed $b$-quark loop.)
Here, $n_f = 5$ and the strong coupling constant $a_s$ is to be taken after decoupling the top quark in the $m_t \rightarrow \infty$ limit.
As a consequence of the RG invariance of the above effective interaction Lagrangian, the RG equation for $C_w(a_s,\mu/m_t)$ in the Larin's scheme reads: %
\begin{eqnarray}
\label{eq:CwRGE}
\mu^2\frac{\mathrm{d}}{\mathrm{d} \mu^2} C_w(a_s, \mu/m_t) 
&=& \mu^2\frac{\partial}{\partial \mu^2} C_w(a_s, \mu/m_t)  \,+\, \beta\, a_s\frac{\partial}{\partial a_s} C_w(a_s, \mu/m_t)  \nonumber\\
&=& \frac{\bar{\gamma}_{s}}{n_f} - \bar{\gamma}_{s} \,C_w(a_s, \mu/m_t)\, ,
%\nonumber\\
%&=& \frac{a_s}{2}\, \gFJ - \bar{\gamma}_{s} \,C_w(a_s, \mu/m_t)\,,
\end{eqnarray} %
with $\bar{\gamma}_{s}$ given in eq.~(\ref{eq:gs5}).
Alternatively, if one takes $Z_s$ (and accordingly $Z_{ps}$) simply in the $\MSbar$ scheme, then the anomalous dimension in the second line of the eq.~(\ref{eq:CwRGE}) will be changed from $\bar{\gamma}_{s}$ to $\gamma^{ms}_{s}$ given in eq.~(\ref{eq:gms5}) and consequently one arrives at the $C_w(a_s,\mu/m_t)$ defined in the $\MSbar$ scheme.
It is straightforward to see that by introducing a particular leading constant term $-1/n_f$, the RG equation for $C_t \equiv -1/n_f + C_w$ takes a more usual form, see, e.g.~ref.~\cite{Chetyrkin:1993jm,Ju:2021lah}.
%It is very attempting to equal the $\bar{\psi}^{B}_{b} \, \gamma^{\mu}\gamma_5 \, \psi^{B}_b$ as simply $\frac{1}{n_f} \big[ J^{\mu}_{5,s} \big]_B$, and consequently the third generation terms in eq.~(\ref{eq:Leff}) can be organised into $a_t \, (-\frac{1}{n_f} + C_w) \big[ J^{\mu}_{5,s} \big]_R = a_t \, C_t \big[ J^{\mu}_{5,s} \big]_R$; however, the aforementioned operator-level identification is not very proper~\cite{Chen:2022aqw}.
Subsequently the solution of the RG equation for the so-defined $C_t(a_s, \mu/m_t)$ in the Larin's scheme can be cast into the following well-known compact form:
\begin{eqnarray}
\label{eq:CwRGEsol}
C_t(a_s(\mu), \mu/m_t) &=& C_t(a_s(m_t), 1) \, \hat{P}\mathrm{exp}\Big(\int_{a_s(m_t)}^{a_s(\mu)} \frac{-\bar{\gamma}_s(a_s)}{\beta(a_s)} \frac{\mathrm{d}a_s}{a_s} \Big)\,,
\end{eqnarray} %
with $C_t(a_s(m_t), 1)$ encoding the integration constant at boundary $\mu=m_t$. 
Obviously, the product $\hat{P}\mathrm{exp}\big(\int_{a_s(m_t)}^{a_s(\mu)} \frac{-\bar{\gamma}_s(a_s)}{\beta(a_s)} \frac{\mathrm{d}a_s}{a_s} \big) \,Z_s\, \big[\partial_{\mu} J^{\mu}_{5,s} \big]_{B}$ is RG-invariant, as this is how the RG equation (\ref{eq:CwRGE}) was determined at the first place.
Replacing $\bar{\gamma}_s$ by $\gamma^{ms}_s$ in the above expression leads to the Wilson coefficient in the $\MSbar$ scheme, and similarly by changing to $\gamma^{ms}_s - \gamma^{ms}_{ns}$ one arrives at the one in Chetyrkin's scheme~\cite{Chetyrkin:1993hk}.
%The RG-invariant integration constant C_t(a_s(m_t), 1) can be seen as effectively the value of the Wilson-coefficient in the aforementioned RG-invariant Chetyrkin-Luescher-Weisz scheme.
The terms featuring $m_t$-logarithms in the perturbative solution for $C_w(a_s,\mu/m_t)$ in Larin's scheme up to $\mathcal{O}(\alpha_s^5)$ can be found in the supplementary material, as well as the finite transformations to the results in $\MSbar$ and Chetyrkin's scheme. (In particular, the result for $\gamma^{ms}_{ns}$ at $\mathcal{O}(\alpha_s^5)$ can be found there.)%
%
%To fully fix all the logarithmic terms of $C_w(a_s,\mu/m_t)$ at $\mathcal{O}(\alpha_s^5)$, one needs the full knowledge of its expression up to $\mathcal{O}(\alpha_s^4)$.
We note that the coefficients of the logarithmic terms of $C_w(a_s,\mu/m_t)$ at higher perturbative orders generally involve the logarithm-independent constant terms of lower perturbative orders, which are typically fixed by matching to an explicit diagrammatic calculation in the full theory in the large $m_t$ limit.
%Since the constant terms clearly depend on the choice of the particular renormalization condition in use, the logarithmic terms will thus become dependent on this as well, albeit starting from a relatively higher perturbative order.
The results for these constant terms were previously determined in refs.~\cite{Chetyrkin:1993jm,Chetyrkin:1993ug,Larin:1993ju,Larin:1994va} up to $\mathcal{O}(a_s^3)$ in Chetyrkin's scheme, and further extended to $\mathcal{O}(a_s^4)$ in the calculation done in ref.~\cite{Baikov:2012er} and published in ref.~\cite{Rittinger:2012the}~(with numerical values of the color factors inserted).
The $\mathcal{O}(a_s^3)$ results in Larin's scheme were provided in refs.~\cite{Ju:2021lah,Chen:2021rft}. 
%One can also solve the RG equation~(\ref{eq:CwRGE}) without using a truncated perturbative ansatz, e.g.~numerically with a given boundary or initial condition, such as the value inferred from a direct experimental measurement at a fixed energy scale, then the resummation would be achieved automatically in the resulting solution.
Finally, we note that if one had chosen the aforementioned scheme with the $Z_{ext}(a_s)$ transformation factor, the corresponding Wilson coefficient $C_t$ would have no more net scale dependence and contain no more explicit logarithm in $m_t$, once expanded in $a_s(m_t)$.
~\\

To summarize, we have determined the $\overline{\mathrm{MS}}$ renormalization constant $Z_{s}^{ms}$ of the singlet axial-current operator in dimensional regularization at $\mathcal{O}(\alpha_s^5)$ in QCD, using a formula derived from the all-order axial anomaly equation combined with the recently proved equality $Z_{F\tilde{F}}=Z_{a_s}$.
%As a preliminary application, we have revealed explicitly the logarithmic part of the Wilson coefficient $C_w(a_s,\mu/m_t)$ at $\mathcal{O}(\alpha_s^5)$ up to one unknown parameter to be fixed by an explicit four-loop calculation.
With the recipe described in this letter, one can efficiently obtain the $\mathcal{O}(a_s^{N+1})$ result for $Z_{s}^{ms}$ from just an $N$-loop calculation.
On the other hand, for $Z_s^f$ at $\mathcal{O}(\alpha_s^5)$, it seems that one still has to perform an explicit five-loop calculation. 
However, once such a computation becomes technically feasible, then the result for $Z_s^{ms}$ at $\mathcal{O}(\alpha_s^6)$ would be immediately within reach. 
It is worthy to emphasize that $Z_s^f$ is, however, not really needed as far as the task of performing resummation of the non-decoupling heavy-quark-mass logarithms for axial quark form factors is concerned, where only non-anomalous non-singlet matrix elements are involved.
~\\

\section*{Acknowledgements}

The work of L.C. and M.C. was supported by the Deutsche Forschungsgemeinschaft under grant 396021762 -- TRR 257.

\bibliography{Zms5} 
\bibliographystyle{utphysM}
% ********** Ending **********
\end{document}